\documentclass[pss]{wiley2sp} 
\usepackage{amsmath}

\usepackage{amssymb}

\newcommand{\angstrom}{\text{\normalfont\AA}}
\mathchardef\mhyphen="2D

\begin{document}

\title{Influence of vibrations on electron transport through nanoscale
  contacts}

\titlerunning{Electron-vibration interactions in charge transport through nanoscale contacts} 

\author{%
  Marius B\"urkle\textsuperscript{\Ast,\textsf{\bfseries 1},$\dagger$},
  Janne K. Viljas\textsuperscript{\textsf{\bfseries 2,3}},
  Thomas J. Hellmuth\textsuperscript{\textsf{\bfseries 1}},  
  Elke Scheer\textsuperscript{\textsf{\bfseries 4}},   
  Florian Weigend\textsuperscript{\textsf{\bfseries 5}},
  Gerd Sch\"on\textsuperscript{\textsf{\bfseries 1,5}}, and
  Fabian Pauly\textsuperscript{\Ast,\textsf{\bfseries 4}}
}

\authorrunning{M. B\"urkle et al.}

\mail{e-mail \textsf{marius.buerkle@aist.go.jp}, email
  \textsf{fabian.pauly@uni-konstanz.de}}

\institute{%
  \textsuperscript{1}\,Institut f\"ur Theoretische
  Festk\"orperphysik, Karlsruhe Institute of Technology, 76131 Karlsruhe,
  Germany\\ \textsuperscript{2}\,Low Temperature Laboratory, Aalto University,
  P.O. Box 15100, 00076 AALTO, Finland\\ \textsuperscript{3}\,Department of
  Physics, P.O. Box 3000, 90014 University of Oulu,
  Finland\\ \textsuperscript{4}\,Department of Physics, University of
  Konstanz, 78457 Konstanz, Germany\\ \textsuperscript{5}\,Institut f\"ur
  Nanotechnologie, Karlsruhe Institute of Technology, 76344
  Eggenstein-Leopoldshafen, Germany\\ \textsuperscript{$\dagger$}\,Present
  address: Nanosystem Research Institute (NRI) "RICS," National Institute of
  Advanced Industrial Science and Technology (AIST), Umezono 1-1-1, Tsukuba
  Central 2, Tsukuba, Ibaraki 305-8568, Japan }

\received{XXXX, revised XXXX, accepted XXXX} 
\published{XXXX} 


\abstract{%

\abstcol{In this article we present a novel semi-analytical approach to
  calculate first-order electron-vibration coupling constants within the
  framework of density functional theory. It combines analytical expressions
  for the first-order derivative of the Kohn-Sham operator with respect to
  nuclear displacements with coupled-perturbed Kohn-Sham theory to determine
  the derivative of the electronic density matrix. This allows us to
  efficiently compute accurate electron-vibration coupling constants.}{We
  apply our approach to describe inelastic electron tunneling spectra of
  metallic and molecular junctions. A gold junction bridged by an atomic chain
  is used to validate the developed method, reproducing established
  experimental and theoretical results. For octanedithiol and octanediamine
  single-molecule junctions we discuss the influence of the anchoring group
  and mechanical stretching on the inelastic electron tunneling spectra.}}

\maketitle   

\section{Introduction}
The advances in nanoscience over the past couple of decades have made it
possible to probe charge transport in nanoscale systems down to the
single-molecule and single-atom level
\cite{Agrait2003,Cuevas2010,Ratner2013}. With such measurements becoming
increasingly routine, less explored effects such as self-heating in
nanojunctions due to current flow move into the focus of research
\cite{Galperin2007,Ioffe2008,Ward2011,Lee2013}. In this context,
electron-vibration (EV) interactions play a crucial role in dissipating the
heat of electrons by transferring it to the vibrational degrees of
freedom. Beside this, inelastic processes can lead to discernible signatures
in transport quantities that can be exploited to characterize nanoscale
conductors. This latter aspect is the subject of the present paper.

The EV interaction can significantly influence transport properties. Thus it
may lead to phonon drag \cite{Jonson1980,Jonson1990,Ziman2001}, thermally
activated transport in long molecular wires \cite{Luo2011}, current saturation
at a high applied voltage bias \cite{Yao2000,Collins2001}, or electromigration
as well as junction breakdown due to current-induced forces and local heating
\cite{Park1999,Smit2004,Taychatanapat2007,Heersche2007,Schirm2013}. Inelastic
EV scattering is also used to characterize molecular and atomic junctions
spectroscopically. Depending on the conductance regime, the techniques are
either called inelastic electron tunneling (IET) spectroscopy
\cite{Stipe1998,Smit2002,Djukic2005} or point contact spectroscopy
\cite{Agrait2003,Agrait2002}. In both cases, however, the second derivative
$\mbox{d}^{2}I/\mbox{d}V^{2}$ of the electrical current $I$ with respect to
the applied bias voltage $V$ is measured. For simplicity we will refer to both
techniques as IET spectroscopy in the following.

In the IET spectroscopy, which is the central subject of this study, the
excitation of a vibrational mode inside the junction gives rise to a
characteristic signature in the current-voltage characteristics. Especially
for molecular junctions, it has become an important tool to identify the
molecule that bridges two metal electrodes and to determine the precise
contact geometry
\cite{Kushmerick2004,Wang2004,Paulsson2006,Kiguchi2008,Arroyo2010,Kim2011}. For
the theoretical description of the EV interaction in nanoscale conductors,
various different approaches exist, which can deal with the whole regime from
weak to intermediate to strong EV couplings
\cite{Galperin2007,Caroli1972,Lorente2001,Koch2005,Galperin2006}.  While model
calculations are able to capture experimentally observed effects on a
qualitative level, a fully atomistic theory is necessary for their
material-dependent, quantitative interpretation. Density functional theory
(DFT) provides such an atomistic electronic structure description for
molecular, solid-state, and hybrid systems. Despite the fact that DFT is not a
quasi-particle method, it is often used to calculate transport properties of
nanoscale junctions in combination with non-equilibrium Green's function
(NEGF) techniques. Approximate DFT tends to underestimate the band gap of bulk
semiconductors or insulators as well as the gap between the highest occupied
and lowest unoccupied molecular orbital of molecular systems. Typically, this
leads to an overestimation of the conductance of molecular junctions. To
compensate for these shortcomings, a ``self-energy corrected'' DFT scheme has
been developed to obtain more accurate quasiparticle energies \cite{Quek2007}
or even time-consuming atomistic quasiparticle methods were employed within
the GW approximation \cite{Strange2011}. For metallic systems, including
metallic atomic contacts, where the ``band gap problem'' of DFT does not
arise, the quantum transport calculations using the combination of DFT and
NEGF (DFT+NEGF) show often good agreement with experiment
\cite{Frederiksen2007b}.

Quantities derived from the total ground-state energy like bond lengths,
binding energies and vibrational spectra are commonly described reliably for
molecular and solid-state systems within the DFT schemes using the local
density approximation (LDA) or generalized gradient approximation (GGA)
\cite{Koch2001,Martin2008,Baroni2001}.  Due to the reasonable compromise
between computational cost and accuracy, DFT+NEGF has become the standard
method for the atomistic, first-principles modeling of transport through
nanoscale devices. The DFT+NEGF approach is very compelling, when inelastic
corrections to the current due to the EV interaction are of interest, since it
allows for the consistent treatment of the whole system
\cite{Frederiksen2007b}: The electronic structure, the vibrational modes, as
well as their coupling can all be described within the same method. In
DFT+NEGF the EV coupling is usually treated in the weak limit either by means
of a lowest-order expansion (LOE) or the self-consistent Born approximation
\cite{Frederiksen2007b,Asai2004,Viljas2005,McEniry2008}.

In this work we extend our cluster-based approach for determining elastic
quantum transport to include the inelastic corrections at the level of the LOE
in the EV coupling. For a detailed discussion of the LOE and of the
cluster-based transport approach, we refer to our previous works in
Refs.~\cite{Viljas2005,Pauly2008}. We do not pursue the direction of
``self-energy corrected'' DFT or atomistic quasiparticle electronic structure
methods here, but use the benefit of DFT to describe the coupled system of
electrons and phonons in atomic and molecular junctions within a single,
unified atomistic approach. We focus particularly on the calculation of the EV
coupling constants within the framework of DFT. We have implemented a scheme
that computes the EV couplings, similar to the phonon modes, using density
functional perturbation theory (DFPT). The use of a Gaussian basis allows us
to calculate the required matrix elements semi-analytically. In this way, we
avoid finite differences, increase the computational efficiency, and prevent
numerical instabilities especially for the low-frequency modes
\cite{SchaeferIII1986}. We test the newly developed method at monovalent gold
(Au) atomic junctions and reproduce literature results for a well-studied
atomic chain configuration. Finally, we discuss the inelastic signals in
molecular junctions. Elaborating on theoretical aspects of our previous work
in Ref.~\cite{Kim2011}, we study the IET spectra of octane-based
single-molecule junctions with thiol and amine anchors. We focus especially on
vibrations localized on the octane molecule, which do not involve electrode
atoms.

According to the work program, this paper is organized as follows. In Section
\ref{sec:Method} we introduce the theoretical methodology, define the
theoretical model of the nanojunction in Subsection
\ref{sub:Definition-system}, show how the electronic and vibrational structure
is obtained from DFT together with the EV couplings in Subsection
\ref{sub:System-within-DFT}, and sketch the LOE to calculate the inelastic
corrections to the current in Subsection \ref{sub:Electrical-current-ev}. In
section \ref{sec:Results} we show applications of our method. We validate our
approach in Subsection \ref{sub:Gold-atomic-contacts} by analyzing a gold
contact in an atomic chain configuration, before we discuss the results for
the octane-based junctions in Subsection \ref{sub:Octane}. Finally, we
conclude with a summary of our results in Section \ref{sec:Conclusions}.

\section{Method\label{sec:Method}}
\subsection{Definition of the system\label{sub:Definition-system}}
We model the nanoscale junctions as a central device region, containing the
atomic or molecular system of interest and parts of the electrodes, which is
connected to semi-infinite, crystalline electrodes to the left and right. The
``dynamical region'' (DR), where atoms can move and vibrations are considered,
is usually identical to the device part, but can also be restricted to a
smaller subset of atoms. At the effective single-particle level, the
Hamiltonian of the coupled system of electrons and vibrations is given by
\cite{Viljas2005}
\begin{equation}
\hat{H}=\hat{H}^{\text{e}}+\hat{H}^{\text{v}}+\hat{H}^{\text{ev}},\label{eq:fullH}
\end{equation}
where the first term
\begin{equation}
\hat{H}^{\text{e}}=\sum_{\mu\nu}\hat{d}_{\mu}^{\dagger}H_{\mu\nu}^{\text{e}}\hat{d}_{\nu}\label{eq:He_2q}
\end{equation}
describes the electronic
system. $H_{\mu\nu}^{\text{e}}=\bigl\langle\mu\bigl|\hat{H}^{\text{e}}_{1}\bigr|\nu\bigr\rangle$
are matrix elements of the single-particle Hamiltonian, represented in first
quantization, in the nonorthogonal, atomic orbital basis $\{ |\mu\rangle \} $,
and $\hat{d}_{\mu}^{\dagger}$ $(\hat{d}_{\mu})$ is the electron creation
(annihilation) operator in that basis, satisfying the anticommutation relation
$\{ \hat{d}_{\mu},\hat{d}_{\nu}^{\dagger} \} =(S^{-1})_{\mu\nu}$. In the
expression $(S^{-1})_{\mu\nu}$ is the inverse of the overlap matrix
$S_{\mu\nu}=\langle\mu | \nu\rangle$. The second term is the Hamiltonian of
the vibrations in the harmonic approximation, given by
\begin{equation}
\hat{H}^{\text{v}}=\sum_{\alpha}\hbar\omega_{\alpha}\hat{b}_{\alpha}^{\dagger}\hat{b}_{\alpha}.
\end{equation}
Here, $\omega_{\alpha}$ is the frequency of the vibrational mode $\alpha$ and
$\hat{b}_{\alpha}^{\dagger}$ $(\hat{b}_{\alpha})$ is the corresponding phonon
creation (annihilation) operator, satisfying the commutation relation
$[\hat{b}_{\alpha},\hat{b}_{\beta}^{\dagger}]=\delta_{\alpha\beta}$.  The
phonon frequencies $\omega_{\alpha}$ are obtained from the eigenvalue problem
\begin{equation}
\boldsymbol{\mathcal{D}\mathcal{C}}^{\alpha}=\omega_{\alpha}^{2}\boldsymbol{\mathcal{C}}^{\alpha},\label{eq:EigvalD}
\end{equation}
with the dynamical matrix
\begin{equation}
\mathcal{D}_{\chi\xi}=\dfrac{1}{\sqrt{M_{k}M_{l}}}\mathcal{H}_{\chi\xi}.\label{eq:dynamicalmatrix}
\end{equation}
Here, $\chi=(k, u)$ and $\xi=(l,v)$ are shorthand notations that refer both to
the displacements of atoms $k,l$ from the equilibrium values of the positions
$\vec{R}_k,\vec{R}_l$ along the Cartesian components $R_{k,u},R_{l,v}$ with
$u,v=\text{x},\text{y},\text{z}$ as well as the index pairs $(k,u)$ and
$(l,v)$ themselves. The matrix
$\mathcal{H}_{\chi\xi}=\mbox{d}^{2}E_{\text{tot}}/\mbox{d}\chi \mbox{d}\xi$ is
the Hessian of the total energy $E_{\text{tot}}$, and $M_k,M_l$ are atomic
masses. The last term in the Hamiltonian
\begin{equation}
\hat{H}^{\text{ev}}=\sum_{\mu\nu}\sum_{\alpha}\hat{d}_{\mu}^{\dagger}\lambda_{\mu\nu}^{\alpha}\hat{d}_{\nu}(\hat{b}_{\alpha}^{\dagger}+\hat{b}_{\alpha})
\end{equation}
describes the first-order EV interaction. The EV coupling constants are
given as
\begin{equation}
\lambda_{\mu\nu}^{\alpha}=\left(\dfrac{\hbar}{2\omega_{\alpha}}\right)^{1/2}\sum_{\chi}\bigl\langle\mu\bigl|\dfrac{\mbox{d}\hat{H}^{\text{e}}_1}{\mbox{d}\chi}\bigr|\nu\bigr\rangle \mathcal{A}_{\chi}^{\alpha},\label{eq:evconst}
\end{equation}
where $\chi=(k,u)$ and
$\mathcal{A}_{\chi}^{\alpha}=\mathcal{C}_{\chi}^{\alpha}/\sqrt{M_k}$ are the
mass-normalized normal modes, obtained from the eigenvectors
$\mathcal{C}_{\chi}^{\alpha}$ of the dynamical matrix in
Eq.~(\ref{eq:EigvalD}).

\subsection{Description of the system within density functional theory\label{sub:System-within-DFT}}

All parameters entering the Hamiltonian in Eq.~(\ref{eq:fullH}) are obtained
in the framework of DFT \cite{Kohn1965}. The basic idea of DFT is to find
variationally the electron density, which delivers the lowest total energy
$E_\text{tot}$, and hence the ground-state energy of the studied many-body
system \cite{Hohenberg1964}. Most of the practical implementations of DFT are
based on the Kohn-Sham (KS) scheme, which maps the interacting many-body
problem onto an effective non-interacting single-particle problem.  We will
make no distinction and simply refer to this ``KS DFT'' as ``DFT'' from here
on. A detailed discussion of DFT can be found in the extensive literature
\cite{Koch2001,Martin2008,Parr1995,Fiolhais2003,Capelle2006,Marques2006}.
Here we will restrict ourselves to the formulas and relations that are
relevant for the present discussion. All of our calculations are based on the
DFT implementation in the quantum chemistry package TURBOMOLE
\cite{Turbomole63}, which uses real Gaussian atomic orbital basis functions.

Finding the ground-state energy in DFT requires the solution of the so-called
KS equations. In the linear combination of atomic orbitals ansatz, the KS
orbitals are expanded in a finite set of basis functions
$\{\langle\vec{r}|\mu\rangle=\phi_{\mu}(\vec{r})\}$.  The resulting equations
are solved self-consistently and are given by
\begin{equation}
\sum_{\nu=1}^{N_{\text{b}}}\left(H_{\mu\nu}^{\text{e}}-\epsilon_{i}S_{\mu\nu}\right)c_{\nu i}=0,\label{eq:ksMatrixEq}
\end{equation}
where $N_{\text{b}}$ is the number of basis functions, $\epsilon_{i}$ is the
energy of molecular orbital $i$, $c_{\nu i}$ are the molecular orbital
expansion coefficients, and $S_{\mu\nu}$ is the overlap matrix introduced
above. The matrix elements of the single-particle Hamiltonian in first
quantization 
$H_{\mu\nu}^{\text{e}}$ are those of the KS ``Fock'' operator
\begin{equation}
\hat{H}^{\text{e}}_1=\hat{h}_1+\hat{J}_1+\hat{V}^{\text{xc}}_1.\label{eq:ks-operator}
\end{equation}
For a system of $N_{\text{e}}$ electrons and $N_{\text{n}}$ nuclei at
positions $\vec{R}_{k}$, the first term of $\hat{H}^{\text{e}}_1$ is given by
the one-electron operator
\begin{equation}
\hat{h}_1=\int\mbox{d}^{3}r\vert\vec{r}\rangle\left[-\dfrac{\hbar^{2}}{2m_{\text{e}}}\nabla^{2}+\sum_{k=1}^{N_{\text{n}}}V_{k}(\vec{r})\right]\langle\vec{r}\vert,
\end{equation}
where the electron mass is $m_{\text{e}}$ and the electron-nucleus interaction
is
$V_{k}(\vec{r})=-e^{2}Z_{k}/(4\pi\epsilon_0\left|\vec{r}-\vec{R}_{k}\right|)$
with the elementary charge $e=|e|$, the vacuum permittivity $\epsilon_0$, and
the atomic number $Z_{k}$ of the $k$-th atom. The second term is the Coulomb
operator
\begin{equation}
\hat{J}_1=\dfrac{e^2}{4\pi\epsilon_0}\int\mbox{d}^{3}r\vert\vec{r}\rangle\int\mbox{d}^{3}r^{\prime}\varrho(\vec{r}^{\prime})\dfrac{1}{\left|\vec{r}-\vec{r}^{\prime}\right|}\langle\vec{r}\vert,
\end{equation}
with the ground-state density
\begin{equation}
\varrho(\vec{r})=\sum_{\mu\nu}\phi_{\mu}(\vec{r})P_{\mu\nu}\phi_{\nu}(\vec{r})
\end{equation}
and the closed-shell density matrix
\begin{equation}
P_{\mu\nu}=2\sum_{i=1}^{N_{\text{e}}/2}c_{\mu i}c_{\nu i}.\label{eq:Pmunu}
\end{equation}
The last term in Eq.~(\ref{eq:ks-operator}) is the exchange-correlation
operator
\begin{equation}
\hat{V}^{\text{xc}}_1=\int\mbox{d}^{3}r\vert\vec{r}\rangle
V^{\text{xc}}([\varrho];\vec{r})\langle\vec{r}\vert,
\end{equation}
which is defined through the functional derivative of the exchange correlation
energy with respect to the charge density,
$V^{\text{xc}}([\varrho];\vec{r})=\delta
E^{\text{xc}}/\delta\varrho(\text{\ensuremath{\vec{r}})}$.  The precise form
of the exchange-correlation energy depends on the choice of the functional
$F([\varrho];\vec{r})$
\begin{equation}
E^{\text{xc}}=\int\mbox{d}^{3}r F([\varrho];\vec{r}).
\end{equation}
With these relations the electronic Hamiltonian $\hat{H}^{\text{e}}$ in
Eq.~(\ref{eq:fullH}) is determined.

To obtain the parameters of the remaining terms $\hat{H}^{\text{v}}$ and
$\hat{H}^{\text{ev}}$ in Eq.~(\ref{eq:fullH}), we need an expression for the
energy. The total DFT ground-state energy is given by
\begin{align}
E_{\text{tot}}&=
\sum_{\mu\nu}P_{\mu\nu}h_{\mu\nu}+\dfrac{1}{2}\sum_{\mu\nu\sigma\kappa}P_{\mu\nu}P_{\sigma\kappa}\bigl(\mu\nu\bigl|\sigma\kappa\bigr)\nonumber
\\ & +E^{\text{xc}}+V^{\text{nn}}.
\end{align}
Here, $h_{\mu\nu}=\langle\mu|\hat{h}_1|\nu\rangle$,
\begin{align}
\bigl(\mu\nu\bigl|\sigma\kappa\bigr)&=\dfrac{e^2}{4\pi\epsilon_0}\int\mbox{d}^{3}r\int\mbox{d}^{3}r^{\prime}\phi_{\mu}(\vec{r})\phi_{\nu}(\vec{r})\nonumber
\\ & \times
\frac{1}{\left|\vec{r}-\vec{r}^{\prime}\right|}\phi_{\sigma}(\vec{r}^{\prime})\phi_{\kappa}(\vec{r}^{\prime})
\label{eq:totalEnergy}
\end{align}
are four-center two-electron Coulomb integrals over Gaussian basis functions,
and the nuclear repulsion energy
$V^{\text{nn}}=\sum_{k=1}^{N_{\text{n}}}\sum_{j>k}^{N_{\text{n}}}e^{2}Z_{k}Z_{j}/(4\pi\epsilon_0|\vec{R}_{k}-\vec{R}_{j}|)$
is given by the last term. The total energy, Eq.~(\ref{eq:totalEnergy}),
depends explicitly on the nuclear coordinates $\vec{R}_{k}$ and on the
molecular orbital expansion coefficients $c_{\mu i}$. Moreover, via
Eq.~(\ref{eq:ksMatrixEq}), the $c_{\mu i}$ depend also on the nuclear
coordinates.

To calculate the vibrational modes in the harmonic approximation, we need the
second derivatives of $E_{\text{tot}}$ with respect to the nuclear displacements $\chi$ and
$\xi$. They are given by \cite{Deglmann2002}
\begin{align}
\dfrac{\mbox{d}^{2}E_{\text{tot}}}{\mbox{d}\chi\mbox{d}\text{\ensuremath{\xi}}}&=\dfrac{\partial^{2}E_{\text{tot}}}{\partial\chi\partial\xi}-\sum_{\mu\nu}\Biggl\{
\dfrac{\partial W_{\mu\nu}}{\partial\xi}\dfrac{\partial
  S_{\mu\nu}}{\partial\chi} \nonumber \\ & 
- W_{\mu\nu}\dfrac{\partial^{^{2}}S_{\mu\nu}}{\partial\chi\partial\xi}+\dfrac{\partial^{2}E_{\text{tot}}}{\partial
  P_{\mu\nu}\partial\chi}\dfrac{\partial P_{\mu\nu}}{\partial\xi}\Biggr\}.
\label{eq:dEdchidxi}
\end{align}
Here, we have defined the energy-weighted density matrix
$W_{\mu\nu}=\sum_{i=1}^{N_{\text{e}}/2}c_{\mu i}\epsilon_{i}c_{\nu i}$.
Beside the derivatives of the one- and two-electron integrals,
Eq.~(\ref{eq:dEdchidxi}) contains also first derivatives of $W_{\mu\nu}$ and
$P_{\mu\nu}$ with respect to $\chi$. They are obtained semi-analytically from
DFPT by means of the first-order coupled-perturbed KS equations
\cite{Pople1979,Baroni1987,Johnson1994,Gonze1995}. The calculation of the
vibrational modes is performed with TURBOMOLE's coupled-perturbed KS
implementation in the module ``aoforce'' \cite{Deglmann2002,Deglmann2004}.

To calculate the EV coupling elements we need, in addition to the
mass-normalized normal modes $\mathcal{A}_\chi^\alpha$, the first derivative
of the KS operator with respect to the nuclear displacements
\begin{equation}
\dfrac{\mbox{d}\hat{H}^{\text{e}}_1}{\mbox{d}\chi}=\dfrac{\partial\hat{H}^{\text{e}}_1}{\partial\chi}+\sum_{\mu\nu}\dfrac{\partial\hat{H}^{\text{e}}_1}{\partial
  P_{\mu\nu}}\dfrac{\partial P_{\mu\nu}}{\partial\chi}.
\end{equation}
The corresponding matrix elements
\begin{align}
\bigl\langle\mu\bigr|\dfrac{\mbox{d}\hat{H}^{\text{e}}_1}{\mbox{d}\chi}\bigl|\nu\bigr\rangle&=
\bigl\langle\mu\bigr|\dfrac{\partial\hat{h}_1}{\partial\chi}\bigl|\nu\bigr\rangle+\bigl\langle\mu\bigr|\dfrac{\mbox{d}\hat{J}_1}{\mbox{d}\chi}\bigl|\nu\bigr\rangle
\nonumber \\ &
+\bigl\langle\mu\bigr|\dfrac{\mbox{d}\hat{V}_{1,\text{LDA}}^{\text{xc}}}{\mbox{d}\chi}\bigl|\nu\bigr\rangle
\end{align}
are given by
\begin{align}
\bigl\langle\mu\bigr|\dfrac{\partial\hat{h}_1}{\partial\chi}\bigl|\nu\bigr\rangle&=-\int\mbox{d}^{3}r\biggl\{\left[\dfrac{\partial\phi_{\mu}(\vec{r})}{\partial\xi}\right]V_{k}(\vec{r})\phi_{\nu}(\vec{r})\nonumber
\\ &
+\phi_{\mu}(\vec{r})V_{k}(\vec{r})\left[\dfrac{\partial\phi_{\nu}(\vec{r})}{\partial\zeta}\right]\biggr\}
 \label{eq:dhdchi}
\end{align}
with $\chi=(k,u)$, while $\xi=(k_{\mu},u)$ and $\zeta=(k_{\nu},u)$ refer to
displacements of the center of the basis functions $\phi_{\mu}$ and
$\phi_{\nu}$, respectively, for the same Cartesian component $u$,
\begin{align}
\bigl\langle\mu\bigr|\dfrac{\mbox{d}\hat{J}_1}{\mbox{d}\chi}\bigl|\nu\bigr\rangle&=
\sum_{\sigma\kappa}\biggl\{\left(\mu\nu\bigr|\sigma\kappa\right)\dfrac{\partial
  P_{\sigma\kappa}}{\partial\chi} \nonumber \\ &
+P_{\sigma\kappa}\bigl(\mu\nu\bigr|\dfrac{\partial}{\partial\chi}[\sigma\kappa]\bigr)\biggr\},
\end{align}
and
\begin{align}
\bigl\langle\mu\bigr|\dfrac{\mbox{d}\hat{V}_{1,\text{LDA}}^{\text{xc}}}{\mbox{d}\chi}\bigl|\nu\bigr\rangle&=\int\mbox{d}^{3}r\phi_{\mu}(\vec{r})\phi_{\nu}(\vec{r})\dfrac{\partial^{2}F_{\text{LDA}}(\varrho(\vec{r}))}{\partial^{2}\varrho}
\nonumber \\ &\times\sum_{\sigma\kappa}\biggl\{P_{\sigma\kappa}
\dfrac{\partial}{\partial\chi}\left[\phi_{\sigma}(\vec{r})\phi_{\kappa}(\vec{r})\right] \nonumber \\ &
+\phi_{\sigma}(\vec{r})\phi_{\kappa}(\vec{r})\dfrac{\partial
  P_{\sigma\kappa}}{\partial\chi}\biggr\} .\label{eq:dVxcdchi}
\end{align}
In the one-electron part, Eq.~(\ref{eq:dhdchi}), we used the translational
invariance to rewrite the Hellmann-Feynman-like expression in terms of the
derivatives of the basis functions \cite{Komornicki1977}. For simplicity we
have quoted the case of the LDA for the derivative of the exchange-correlation
operator, as indicated by the subscripts, and discussed closed shell
systems. Yet, our implementation in the development version of TURBOMOLE
handles LDA, GGA, and meta-GGA functionals
in the spin-restricted and spin-unrestricted case.

\subsection{Electrical current including inelastic effects due to
  electron-vibration interactions\label{sub:Electrical-current-ev}} We model
the atomic or molecular junctions through an extended central cluster (ECC)
(see Ref.~\cite{Pauly2008} as well as Figs.~\ref{fig:goldECC_tran} and
\ref{fig:octacestretch}), which contains the narrowest constriction and large
parts of the electrodes. It is subsequently divided into a central (C) region
and the parts belonging to the left (L) and right (R) electrodes. From the
ECC, the information on the C region and its couplings to the L and R parts
are extracted. For the description of the semi-infinite, perfect-crystal
electrodes in the L and R regions we perform separate calculations to
determine their bulk-like electronic structure. We do not discuss them here,
but all the details can be found in Ref.~\cite{Pauly2008}.

Using the LOE in the EV coupling, as developed in Ref.~\cite{Viljas2005}, we
express the current through the C region
\begin{equation}
I=I_{\text{el}}+\delta I_{\text{el}}+I_{\text{inel}}\label{eq:Itot}
\end{equation}
as the sum of the elastic contribution
\begin{equation}
I_{\text{el}}=\dfrac{2e}{h}\int\mbox{d}E\tau(E)[f_{\text{L}}(E)-f_{\text{R}}(E)],\label{eq:Iel0}
\end{equation}
a quasi-elastic correction corresponding to the emission and reabsorption of a
virtual phonon, which does not change the energy of the scattered electron,
\begin{align}
\delta I_{\text{el}}&=
\dfrac{4e}{h}\int\mbox{d}E\mbox{ReTr}[\boldsymbol{\Gamma}_{\text{L}}(E)\boldsymbol{G}^{\text{r}}(E)\boldsymbol{\Sigma}_{\text{ev}}^{\text{r}}(E)\boldsymbol{G}^{\text{r}}(E)\nonumber
  \\ &
  \times\boldsymbol{\Gamma}_{\text{R}}(E)\boldsymbol{G}^{\text{a}}(E)][f_{\text{L}}(E)-f_{\text{R}}(E)],\label{eq:deltaIel0}
\end{align}
and an inelastic correction due to the emission or absorption of a real phonon
by an electron
\begin{align}
  I_{\text{inel}} & = -\mbox{i}\dfrac{2e}{h}\int\mbox{d}E\mbox{Tr}\bigl[
    \boldsymbol{G}^{\text{a}}(E)\boldsymbol{\Gamma}_{\text{L}}(E)\boldsymbol{G}^{\text{r}}(E)
    \nonumber \\ & \times \bigl\{
              [f_{\text{L}}(E)-1]\boldsymbol{\Sigma}_{\text{ev}}^{<}(E)-f_{\text{L}}(E)\boldsymbol{\Sigma}_{\text{ev}}^{>}(E)
              \bigr\} \bigr]. \label{eq:Iinel}
\end{align}
Here, the Fermi function of the lead $X=\text{L},\text{R}$ is given by
$f_{X}(E)=f(E-\mu_{X})$ with $f(E)=1/[\exp(\beta E)+1]$,
$\beta=1/(k_{\text{B}}T)$, the Boltzmann constant $k_{\text{B}}$, and the
temperature $T$. We assume the electrochemical potentials in the L and R
electrodes to be $\mu_\text{L}=E_{\text{F}}+U/2$ and
$\mu_{\text{R}}=E_{\text{F}}-U/2$ with the Fermi energy $E_{\text{F}}$ and the
potential $U=eV$ due to the applied bias.

In Eq.~(\ref{eq:Iel0}),
\begin{align}
\tau(E)&=\mbox{Tr}\left[\boldsymbol{G}^{\text{r}}(E)\boldsymbol{\Gamma}_{\text{R}}(E)\boldsymbol{G}^{\text{a}}(E)\boldsymbol{\Gamma}_{\text{L}}(E)\right]
\nonumber\\
&=\sum_{i}\tau_{i}(E)
\end{align}
is the energy-dependent elastic transmission which can be decomposed into the
contribution of different transmission eigenchannels $i$. With the techniques
of Refs.~\cite{Pauly2008,Paulsson2007,Burkle2012} both the eigenchannel
transmission probability $\tau_{i}(E)$ as well as the corresponding
scattering-state wave-function $\Psi_{i}(\vec{r},E)$ can be
determined. Ignoring the inelastic contributions, the conductance at low
temperatures and in the linear response regime is $G=G_{0}\tau(E_{\text{F}})$
with the quantum of conductance $G_{0}=2e^{2}/h$.

The lowest-order EV self-energies are given by
\cite{Frederiksen2007b,Viljas2005,Hyldgaard1994,Dash2010}
\begin{align}
\boldsymbol{\Sigma}_{\text{ev}}^{\lessgtr}(E)&= \dfrac{\mbox{i}}{2\pi}\sum_{\alpha}\int\mbox{d}E^{\prime}D_{\alpha}^{\lessgtr}(E^{\prime}) \nonumber \\
 & \times\boldsymbol{\lambda}^{\alpha}\boldsymbol{G}^{\lessgtr}(E-E^{\prime})\boldsymbol{\lambda}^{\alpha}
\end{align}
and
\begin{equation}
\boldsymbol{\Sigma}_{\text{ev}}^{\text{r},\text{a}}(E)=\boldsymbol{\Sigma}_{\text{H}}^{\text{r},\text{a}}+\boldsymbol{\Sigma}_{\text{F}}^{\text{r},\text{a}}(E),
\end{equation}
where the two contributions in
$\boldsymbol{\Sigma}_{\text{ev}}^{\text{r},\text{a}}(E)$ are the Hartree
term
\begin{equation}
\boldsymbol{\Sigma}_{\text{H}}^{\text{r},\text{a}}=-\dfrac{\mbox{i}}{2\pi}\sum_{\alpha}D_{\alpha}^{\text{r}}(0)\boldsymbol{\lambda}^{\alpha}\int\mbox{d}E\mbox{Tr}[\boldsymbol{G}^{<}(E)\boldsymbol{\lambda}^{\alpha}]
\end{equation}
and the Fock term
\begin{align}
\boldsymbol{\Sigma}_{\text{F}}^{\text{r},\text{a}}(E) &
=\dfrac{\mbox{i}}{2\text{\ensuremath{\pi}}}\sum_{\alpha}\int\mbox{d}E^{\prime}\bigl[D_{\alpha}^{<}(E^{\prime})\boldsymbol{\lambda}^{\alpha}\boldsymbol{G}^{\text{r},\text{a}}(E-E^{\prime})\boldsymbol{\lambda}^{\alpha}\nonumber
\\ &
+D_{\alpha}^{\text{r},\text{a}}(E^{\prime})\boldsymbol{\lambda}^{\alpha}\boldsymbol{G}^{>}(E-E^{\prime})\boldsymbol{\lambda}^{\alpha}\bigr].\label{eq:sigma_ra_f}
\end{align}
In the above and all the following equations, the summation over $\alpha$ runs
over all vibrations in the DR. For a number of $N_{\text{DR}}$ atoms, this
yields $3N_{\text{DR}}$ modes. Typically, we choose $N_{\text{DR}}$ equal to
the number $N_{\text{C}}$ of atoms in the center, but it can also be smaller.

We note that our Green's function matrices $\boldsymbol{G}^{\lessgtr}$ are
identical to $\boldsymbol{G}^{\pm\mp}$ of Ref.~\cite{Viljas2005}, and the
corresponding self-energies are connected by
$\boldsymbol{\Sigma}^{\lessgtr}=-\boldsymbol{\Sigma}^{\pm\mp}$.  Compared to
Ref. \cite{Frederiksen2007b},
$\boldsymbol{\Sigma}_{\text{ev}}^{\text{r},\text{a}}(E)$ differs by the
Hartree contribution $\boldsymbol{\Sigma}_{\text{H}}^{\text{r},\text{a}}$,
which is disregarded there. In the wide-band limit (WBL) approximation,
introduced further below [see Eqs.~(\ref{eq:IelWBL})-(\ref{eq:IinelWBL})], the
Hartree term yields no contribution to $\mbox{d}^{2}I/\mbox{d}V^{2}$.

The electronic Green's functions are determined through
\begin{equation}
\boldsymbol{G}^{\text{r}}(E)=\left[E\boldsymbol{S}_{\text{CC}}-\boldsymbol{H}^{\text{e}}_{\text{CC}}-\boldsymbol{\Sigma}_{\text{L}}^{\text{r}}(E)-\boldsymbol{\Sigma}_{\text{R}}^{\text{r}}(E)\right]^{-1},
\end{equation}
\begin{equation}
\boldsymbol{G}^{\lessgtr}(E)=\boldsymbol{G}^{\text{r}}(E)\bigl[\boldsymbol{\Sigma}_{\text{L}}^{\lessgtr}(E)+\boldsymbol{\Sigma}_{\text{R}}^{\lessgtr}(E)\bigr]\boldsymbol{G}^{\text{a}}(E),
\end{equation}
and $\boldsymbol{G}^{\text{a}}=(\boldsymbol{G}^{\text{r}})^{\dagger}$. In the LOE
the EV self-energy is not included in $\boldsymbol{G}^{\text{r}}$. The semi-infinite leads are taken into account via the lead self-energies
\begin{align}
\boldsymbol{\Sigma}_{X}^{\text{r}}(E)&=(\boldsymbol{H}^{\text{e}}_{\text{C}X}-E\boldsymbol{S}_{\text{C}X})\nonumber\\ &\times\boldsymbol{g}_{XX}^{\text{r}}(E)(\boldsymbol{H}^{\text{e}}_{X\text{C}}-E\boldsymbol{S}_{X\text{C}}),
\end{align}
\begin{equation}
\boldsymbol{\Sigma}_{X}^{<}(E)=\mbox{i}\boldsymbol{\Gamma}_{X}(E)f_{X}(E),
\end{equation}
\begin{equation}
\boldsymbol{\Sigma}_{X}^{>}(E)=\mbox{i}\boldsymbol{\Gamma}_{X}(E)\left[f_{X}(E)-1\right]
\end{equation}
and linewidth-broadening matrices
\begin{equation}
\boldsymbol{\Gamma}_{X}(E)=-2\mbox{Im}\boldsymbol{\Sigma}_{X}^{\text{r}}(E).
\end{equation}
In the expressions,
$\boldsymbol{g}_{XX}^{\text{r}}(E)=[(E+\mbox{i}\varepsilon)\boldsymbol{S}_{XX}-\boldsymbol{H}^{\text{e}}_{XX}]^{-1}$
is the surface Green's function of the semi-infinite lead
$X=\text{L},\text{R}$ with a small $\varepsilon>0$.

Following Ref.~\cite{Viljas2005}, we approximate the retarded phonon Green's function
by the free propagator
\begin{align}
D_{\alpha}^{\text{r}}(E)\approx d_{\alpha}^{\text{r}}(E)&=
\dfrac{1}{E-E_{\alpha}+\mbox{i}\eta/2} \nonumber \\ &
-\dfrac{1}{E+E_{\alpha}+\mbox{i}\eta/2},
\end{align}
and the lesser and greater phonon Green's functions are expressed in terms of the
non-equilibrium vibrational distribution function $N_{\alpha}(E)$ as
\begin{equation}
D^{<}_{\alpha}(E)=-2\pi\mbox{i}N_{\alpha}(E)\rho_{\alpha}(E),
\end{equation}
\begin{equation}
D^{>}_{\alpha}(E)=-2\pi\mbox{i}[N_{\alpha}(E)+1]\rho_{\alpha}(E).
\end{equation}
Here, $E_{\alpha}=\hbar\omega_{\alpha}$ are the bare vibrational energies. The
vibrational spectral density
$\rho_{\alpha}=-\mbox{Im}D_{\alpha}^{\text{r}}(E)/\pi\approx-\mbox{Im}d_{\alpha}^{\text{r}}(E)/\pi$
is approximated by the imaginary part of the free phonon Green's function
$d_{\alpha}^{\text{r}}(E)$. By keeping the infinitesimal quantity $\eta$ finite,
we approximately account for the finite life-time of the vibrations in the DR due to
the coupling to the electrodes and the environment. The vibrational spectral density
becomes 
\begin{align}
\rho_{\alpha}(E)&=
\dfrac{1}{2\pi}\biggl[\dfrac{\eta}{(E-E_{\alpha})^{2}+\eta^{2}/4} \nonumber
  \\ & -\dfrac{\eta}{(E+E_{\alpha})^{2}+\eta^{2}/4}\biggr],
\end{align}
 and the corresponding non-equilibrium voltage- and temperature-dependent
 vibrational distribution function
\begin{equation}
N_{\alpha}(E)=\dfrac{1}{2}\dfrac{\mbox{Im}\Pi_{\alpha}^{<}(E)-n(E)\eta
  E/E_{\alpha}}{\mbox{Im}\Pi_{\alpha}^{\text{r}}(E)-\eta E/(2E_{\alpha})}
\end{equation}
with the Bose function $n(E)=1/[\exp(\beta E)-1]$ describes the heating and cooling effects in the DR. Here,
\begin{align}
\Pi_{\alpha}^{<}(E)&=
-\dfrac{\mbox{i}}{2\pi}\int\mbox{d}E^{\prime}\mbox{Tr}\bigl[\boldsymbol{\lambda}^{\alpha}\boldsymbol{G}^{<}(E^{\prime}) \nonumber\\ &
  \times \boldsymbol{\lambda}^{\alpha}\boldsymbol{G}^{>}(E^{\prime}-E)\bigr]
\end{align}
and
\begin{align}
\Pi_{\alpha}^{\text{r}}(E)&=
-\dfrac{\mbox{i}}{2\pi}\int\mbox{d}E^{\prime}\mbox{Tr}\bigl[\boldsymbol{\lambda}^{\alpha}\boldsymbol{G}^{<}(E^{\prime})\boldsymbol{\lambda}^{\alpha}\boldsymbol{G}^{\text{a}}(E^{\prime}-E)
  \nonumber \\ &
  +\boldsymbol{\lambda}^{\alpha}\boldsymbol{G}^{\text{r}}(E^{\prime})\boldsymbol{\lambda}^{\alpha}\boldsymbol{G}^{<}(E^{\prime}-E)\bigr]
\end{align}
are the lesser and the retarded phonon self-energies. Using the definitions of
the transport coefficients as given in Ref.~\cite{Viljas2005}
\begin{equation}
\tau(E)=\mbox{Tr}\left[\boldsymbol{G}^{\text{r}}(E)\boldsymbol{\Gamma}_{\text{R}}(E)\boldsymbol{G}^{\text{a}}(E)\boldsymbol{\Gamma}_{\text{L}}(E)\right],\label{eq:tau}
\end{equation}
\begin{align}
T_{\sigma\alpha}^{\text{in}}(E,E^{\prime})&= \mbox{Tr}[\boldsymbol{G}^{\text{r}}(E_{\sigma})\boldsymbol{\Gamma}_{\text{R}}(E_{\sigma})\boldsymbol{G}^{\text{a}}(E_{\sigma})\boldsymbol{\lambda}^{\alpha} \nonumber \\ 
 & \times\boldsymbol{G}^{\text{a}}(E)\boldsymbol{\Gamma}_{\text{L}}(E)\boldsymbol{G}^{\text{r}}(E)\boldsymbol{\lambda}^{\alpha}],
\end{align}
\begin{align}
T_{\sigma\alpha}^{\text{ec}}(E,E^{\prime})&=
2\mbox{ReTr}[\boldsymbol{G}^{\text{r}}(E)\boldsymbol{\Gamma}_{\text{R}}(E)\boldsymbol{G}^{\text{a}}(E) \nonumber
  \\ &
  \times\boldsymbol{\Gamma}_{\text{L}}(E)\boldsymbol{G}^{\text{r}}(E)\boldsymbol{\lambda}^{\alpha}\boldsymbol{G}^{\text{r}}(E_{\sigma})\boldsymbol{\lambda}^{\alpha}],
\end{align}
\begin{align}
T_{\sigma\alpha}^{\text{ec}X}(E,E^{\prime})&=
\mbox{ImTr}[\boldsymbol{G}^{\text{r}}(E)\boldsymbol{\Gamma}_{\text{R}}(E)\boldsymbol{G}^{\text{a}}(E)\boldsymbol{\Gamma}_{\text{L}}(E)\boldsymbol{G}^{\text{r}}(E) \nonumber
  \\ &
  \times \boldsymbol{\lambda}^{\alpha}\boldsymbol{G}^{\text{r}}(E_{\sigma})\boldsymbol{\Gamma}_{X}(E_{\sigma})\boldsymbol{G}^{\text{a}}(E_{\sigma})\boldsymbol{\lambda}^{\alpha}],
\end{align}
\begin{align}
J_{\alpha}^{X}(E) &
=\dfrac{1}{\pi}\int\mbox{d}E^{\prime}\mbox{Re}D_{\alpha}^{\text{r}}(E^{\prime})\mbox{ReTr}[\boldsymbol{G}^{\text{r}}(E)\nonumber
  \\ &
  \times\boldsymbol{\Gamma}_{\text{R}}(E)\boldsymbol{G}^{\text{a}}(E)\boldsymbol{\Gamma}_{\text{L}}(E)\boldsymbol{G}^{\text{r}}(E)\boldsymbol{\lambda}^{\alpha}\nonumber
  \\ &
  \times\boldsymbol{G}^{\text{r}}(E-E^{\prime})\boldsymbol{\Gamma}_{X}(E-E^{\prime})\nonumber
\\ &\times\boldsymbol{G}^{\text{a}}(E-E^{\prime})\boldsymbol{\lambda}^{\alpha}]
  f_{X}(E-E^{\prime}),
\end{align}
\begin{align}
T_{\alpha}^{\text{II}}(E)&= 2\mbox{ReTr}[\boldsymbol{G}^{\text{r}}(E)\boldsymbol{\Gamma}_{\text{R}}(E)\boldsymbol{G}^{\text{a}}(E) \nonumber \\
 & \times\boldsymbol{\Gamma}_{\text{L}}(E)\boldsymbol{G}^{\text{r}}(E)\boldsymbol{\lambda}^{\alpha}],\label{eq:TII}
\end{align}
\begin{align}
J_{\alpha}^{\text{II}X}&=
\dfrac{D_{\alpha}^{\text{r}}(0)}{2\pi}\int\mbox{d}E\mbox{Tr}[\boldsymbol{G}^{\text{r}}(E)\boldsymbol{\Gamma}_{X}(E)
  \nonumber \\ &
  \times\boldsymbol{G}^{\text{a}}(E)\boldsymbol{\lambda}^{\alpha}]f_{X}(E),\label{eq:JIIX}
\end{align}
and the abbreviations $X=\text{L},\text{R}$ and $E_{\sigma}=E+\sigma
E^{\prime}$ with $\sigma=\pm 1$, we can express the current formulas in
Eqs.~(\ref{eq:Iel0})-(\ref{eq:Iinel}) as
\begin{equation}
I_{\text{el}}=\dfrac{2e}{h}\int\mbox{d}E\tau(E)\left[f_{\text{L}}(E)-f_{\text{R}}(E)\right],
\end{equation}
\begin{align}
\delta I_{\text{el}} &
=\dfrac{2e}{h}\int\mbox{d}E\sum_{\alpha}\sum_{\sigma=\pm1}\sigma\int_{0}^{\infty}\mbox{d}E^{\prime}\rho_{\alpha}(E^{\prime})\nonumber
\\ & \times\bigl[ T_{\sigma\alpha}^{\text{ec}}(E,E^{\prime})N_{\alpha}(\sigma
E^{\prime})\nonumber\\ &+T_{\sigma\alpha}^{\text{ecL}}(E,E^{\prime})f_{\text{L}}(E_{\sigma})\nonumber
\\ &
+T_{\sigma\alpha}^{\text{ecR}}(E,E^{\prime})f_{\text{R}}(E_{\sigma})\bigr]
\left[f_{\text{L}}(E)-f_{\text{R}}(E)\right]\nonumber \\ &
-\dfrac{2e}{h}\int\mbox{d}E\sum_{\alpha}\bigl[
J_{\alpha}^{\text{L}}(E)+J_{\alpha}^{\text{R}}(E)\nonumber \\ &
-T_{\alpha}^{\text{II}}(E)\left(J_{\alpha}^{\text{IIL}}+J_{\alpha}^{\text{IIR}}\right)\bigr]
\left[f_{\text{L}}(E)-f_{\text{R}}(E)\right],
\end{align}
\begin{align}
I_{\text{inel}}&=
\dfrac{2e}{h}\int\mbox{d}E\sum_{\alpha}\sum_{\sigma=\pm1}\sigma\int_{0}^{\infty}\mbox{d}E^{\prime}\rho_{\alpha}(E^{\prime})\nonumber
\\ &\times T_{\sigma\alpha}^{\text{in}}(E,E^{\prime}) \bigl\{
N_{\alpha}(\sigma
E^{\prime})f_{\text{L}}(E)\left[1-f_{\text{R}}(E_{\sigma})\right] \nonumber\\&
+ N_{\alpha}(-\sigma
E^{\prime})f_{\text{R}}(E_{\sigma})\left[1-f_{\text{L}}(E)\right]\bigr\}.
\end{align}

In the following calculations we will assume in addition that the
energy-dependent electronic Green's functions are constant around the Fermi
energy $E_{\text{F}}$. In this so-called WBL the transport coefficients, defined in
Eqs.~(\ref{eq:tau})-(\ref{eq:JIIX}), simplify to
\begin{equation}
\tau=\mbox{Tr}\left.\left[\boldsymbol{G}^{\text{r}}\boldsymbol{\Gamma}_{\text{R}}\boldsymbol{G}^{\text{a}}\boldsymbol{\Gamma}_{\text{L}}\right]\right|_{E_{\text{F}}},
\end{equation}
\begin{equation}
T_{\alpha}^{\text{in}}=\mbox{Tr}\left.\left[\boldsymbol{G}^{\text{r}}\boldsymbol{\Gamma}_{\text{R}}\boldsymbol{G}^{\text{a}}\boldsymbol{\lambda}^{\alpha}\boldsymbol{G}^{\text{a}}\boldsymbol{\Gamma}_{\text{L}}\boldsymbol{G}^{\text{r}}\boldsymbol{\lambda}^{\alpha}\right]\right|_{E_{\text{F}}},
\end{equation}
\begin{equation}
T_{\alpha}^{\text{ec}}=2\mbox{ReTr}\left.\left[\boldsymbol{G}^{\text{r}}\boldsymbol{\Gamma}_{\text{R}}\boldsymbol{G}^{\text{a}}\boldsymbol{\Gamma}_{\text{L}}\boldsymbol{G}^{\text{r}}\boldsymbol{\lambda}^{\alpha}\boldsymbol{G}^{\text{r}}\boldsymbol{\lambda}^{\alpha}\right]\right|_{E_{\text{F}}},
\end{equation}
\begin{equation}
T_{\alpha}^{\text{ec}X}
=\left.\mbox{ImTr}\left[\boldsymbol{G}^{\text{r}}\boldsymbol{\Gamma}_{\text{R}}\boldsymbol{G}^{\text{a}}\boldsymbol{\Gamma}_{\text{L}}\boldsymbol{G}^{\text{r}}\boldsymbol{\lambda}^{\alpha}\boldsymbol{G}^{\text{r}}\boldsymbol{\Gamma}_{X}\boldsymbol{G}^{\text{a}}\boldsymbol{\lambda}^{\alpha}\right]\right|_{E_{\text{F}}},
\end{equation}
\begin{equation}
T_{\alpha}^{\text{J}X}=\mbox{ReTr}\left.\left[\boldsymbol{G}^{\text{r}}\boldsymbol{\Gamma}_{\text{R}}\boldsymbol{G}^{\text{a}}\boldsymbol{\Gamma}_{\text{L}}\boldsymbol{G}^{\text{r}}\boldsymbol{\lambda}^{\alpha}\boldsymbol{G}^{\text{r}}\boldsymbol{\Gamma}_{X}\boldsymbol{G}^{\text{a}}\boldsymbol{\lambda}^{\alpha}\right]\right|_{E_{\text{F}}},
\end{equation}
\begin{equation}
T^{\text{II}}_{\alpha}=2\mbox{ReTr}\left.\left[\boldsymbol{G}^{\text{r}}\boldsymbol{\Gamma}_{\text{R}}\boldsymbol{G}^{\text{a}}\boldsymbol{\Gamma}_{\text{L}}\boldsymbol{G}^{\text{r}}\boldsymbol{\lambda}^{\alpha}\right]\right|_{E_{\text{F}}}.
\end{equation}
Not all integrals appearing in the expression for the current do converge
separately in the WBL. However, it is possible to combine them such that they
converge, yielding well-defined results. By doing so, one of the two energy
integrations can be carried out analytically, simplifying the expressions for
the current to
\begin{equation}
I_{\text{el}}=\dfrac{2e}{h}\tau U,\label{eq:IelWBL}
\end{equation}
\begin{align}
\delta
I_{\text{el}}&=\dfrac{2e}{h}\sum_{\alpha}\biggl\{\int_{0}^{\infty}\mbox{d}E\rho_{\alpha}(E)\Bigl\{T_{\alpha}^{\text{ec}}\left[2N_\alpha(E)+1\right]U\nonumber
\\ & + (T_{\alpha}^{\text{ecL}}+T_{\alpha}^{\text{ecR}})[(E-U)n(E-U) \nonumber
  \\ &
  -(E+U)n(E+U)-U]\Bigr\}-\dfrac{1}{\pi}(T_{\alpha}^{\text{JR}}-T_{\alpha}^{\text{JL}})\nonumber\\ &
\times\int\mbox{d}E\mbox{Re}D_{\alpha}^{\text{r}}(E)[En(E)-(E+U)n(E+U)]
\nonumber \\ &
+\dfrac{1}{2}T_{\alpha}^{\text{II}}D_{\alpha}^{\text{r}}(0)\mbox{Tr}[\boldsymbol{P}^{\text{ne}}\boldsymbol{\lambda}^{\alpha}]U\biggl\},\label{eq:dIelWBL}
\end{align}
\begin{align}
I_{\text{inel}}&=
\dfrac{2e}{h}\sum_{\alpha}T_{\alpha}^{\text{in}}\int_{0}^{\infty}\mbox{d}E\rho_{\alpha}(E)\nonumber\\ &\times
\bigl[2N_\alpha(E)U +(E-U)n(E-U) \nonumber \\ &-
  (E+U)n(E+U) \bigr].\label{eq:IinelWBL}
\end{align}
The remaining energy integration over $E$ can be carried out by standard
numerical quadrature, and in Eq.~(\ref{eq:dIelWBL}) the non-equilibrium
density matrix $\boldsymbol{P}^{\text{ne}}=-\mbox{i}\int \mbox{d}E
\boldsymbol{G}^<(E)/\pi$ is approximated by Eq.~(\ref{eq:Pmunu}).

\section{Results and discussion\label{sec:Results}}

In this section, we will apply the methodology of Section \ref{sec:Method} to
study the influence of vibrations on electron transport. Regarding the
computational details, all the calculations are performed with the GGA
exchange-correlation functional BP86 \cite{Becke1988,Perdew1986} and the
module ``ridft'' of TURBOMOLE \cite{Turbomole63}. We use the basis set
def-SV(P) \cite{Schafer1992,Eichkorn1995,Eichkorn1997}, which is of
split-valence quality with polarization functions on all non-hydrogen
atoms. For Au atoms we use an electronic core potential to efficiently deal
with the innermost 60 electrons \cite{Andrae1990}, while our basis sets
explicitly consider all electrons for the rest of the atoms.

\subsection{Gold atomic contacts\label{sub:Gold-atomic-contacts}}

\begin{figure*}[!t]
\includegraphics[width=\textwidth]{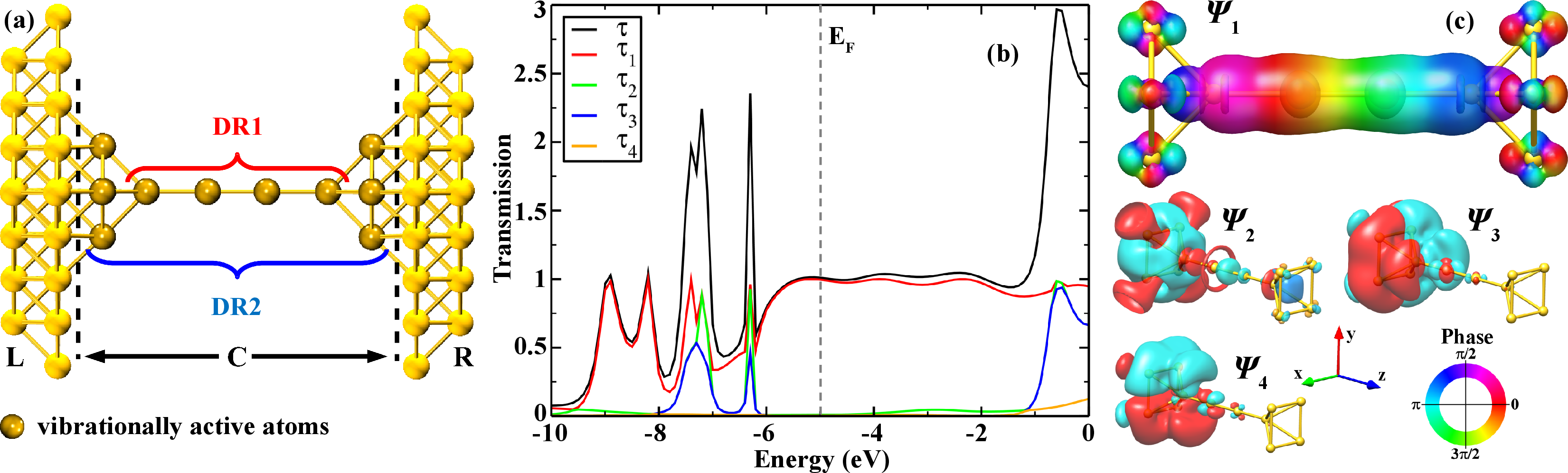}
\caption{(a) Chain of four Au atoms connected to Au electrodes. All atoms in
  the C region have been relaxed. The dynamical regions, where atoms can
  vibrate, are marked as DR1 and DR2. (b) Energy-dependent transmission
  $\tau(E)$ and the four largest transmission probabilities $\tau_{1}(E)$ to
  $\tau_{4}(E)$ of the eigenchannels. The Fermi energy $E_\text{F}$ is
  indicated by a vertical dashed line. (c) Wavefunctions $\Psi_{1}$ to
  $\Psi_{4}$ of the corresponding, left-incoming transmission eigenchannels,
  evaluated at the Fermi energy. The isosurface value is
  0.012~$\angstrom^{-3/2}$ for $\Psi_{1}$ and 0.004~$\angstrom^{-3/2}$ for
  $\Psi_{2}$ to $\Psi_{4}$.\label{fig:goldECC_tran}}
\end{figure*}

To test and validate the transport method, we examine a four-atom gold
chain. It is connected to two Au $\langle 100\rangle$ electrodes, each
consisting of 45 atoms, as shown in Fig.~\ref{fig:goldECC_tran}a. We started
from an ideal geometry, constructed using a lattice constant of
$a=4.08$~\angstrom. Then, the C region, consisting of the four chain atoms and
the closest four atoms of each electrode (see Fig.~\ref{fig:goldECC_tran}a),
was fully relaxed, while the other atoms in the L and R parts were kept fixed
at their ideal face-centered cubic Bravais lattice positions. This system has
already been studied with respect to its elastic conductance
\cite{Mozos2002,Brandbyge2002,Lee2004}, transmission eigenchannel
wavefunctions \cite{Paulsson2007}, and inelastic signatures in the
current-voltage characteristics due to the EV coupling
\cite{Frederiksen2007b}. It therefore serves as an ideal test system for our
newly developed method.

In Fig.~\ref{fig:goldECC_tran}b the elastic transmission $\tau$ and the four
largest transmission eigenchannel probabilities, $\tau_{1}$ to $\tau_{4}$, are
displayed as a function of energy. In agreement with previous studies
\cite{Pauly2008,Mozos2002,Brandbyge2002,Lee2004} we find that $\tau(E)$ is
roughly constant around the Fermi energy $E_{\text{F}}=-5.0$~eV and close to 1 for
energies between $-5.5$ and $-2.0$ eV. The peaks occurring between $-8.0$ and
$-6.0$ eV are due to Au $d$ states. Consistent with experimental results
\cite{Ohnishi1998,Yanson1998,Smit2001}, we obtain a conductance of
$G=1.01G_0$. The channel decomposition of the transmission shows that at
$E_{\text{F}}$ the transport is carried by one almost completely transparent
channel with $\tau_{1}=0.996$. The corresponding, left-incoming eigenchannel
wavefunction $\Psi_1$ in Fig.~\ref{fig:goldECC_tran}c is evaluated at
$E_{\text{F}}$, using the procedure described in Ref.~\cite{Burkle2012}. It
possesses rotational symmetry in the transport direction along the chain,
which is assumed to coincide with the $z$ axis. Due to this $\sigma$ symmetry,
$\Psi_1$ is mainly formed from the $s$ and $p_z$ valence orbitals of the Au
atoms. The phase-factor is color-coded onto the isosurface of the
wavefunction. We observe that it changes continuously along the transport
direction, as expected for a propagating wave. The next three transmission
channels with $\tau_{2}=0.009$, $\tau_{3}=0.003$ and $\tau_{4}=0.003$ arise
from tails of transmission resonances of $d$ states around 1.5 eV below
$E_{\text{F}}$. These eigenchannels constitute evanescent waves, decaying along the
chain. For this reason, we can choose the phase-factors such that the
wave-functions $\Psi_2$ to $\Psi_4$ are purely real in that
region. Figure~\ref{fig:goldECC_tran}c visualizes the $d$ character of the
wavefunctions on the Au chain. The wavefunction $\Psi_2$ of the second channel
is mainly attributed to $d_{3\text{z}^{2}\text{-r}^{2}}$ states with $\sigma$
symmetry along the transport direction.  Channels three and four of $\pi$
symmetry are almost degenerate at $E_{\text{F}}$ and their wavefunctions are
formed from Au chain $d_{\text{xz}}$ and $d_{\text{yz}}$ orbitals,
respectively. The channels involving the remaining two $d$ states,
$d_{\text{xy}}$ and $d_{\text{x}^{2}\text{-y}^{2}}$, have a much smaller
transmission and are not shown here.

\begin{figure*}[t]
\centering{}\includegraphics[width=1.8\columnwidth]{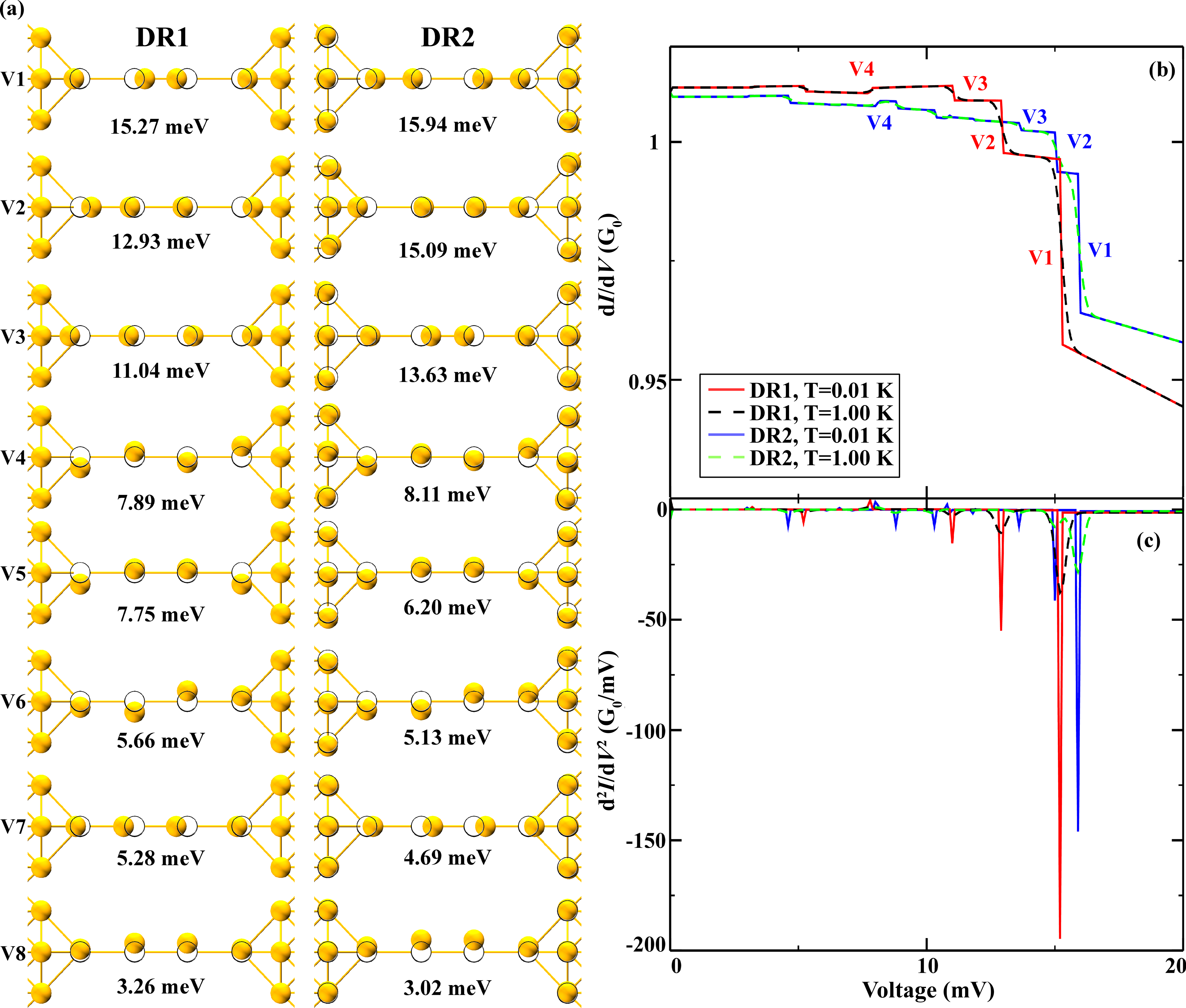}\caption{(a)
  Vibrational modes for DR1 and DR2 with the dynamical regions defined as
  shown in Fig.~\ref{fig:goldECC_tran}a. For DR2 those modes are displayed,
  which are mainly localized on the chain and resemble those of DR1. (b)
  Differential conductance as a function of voltage for DR1 and DR2 at the
  temperatures of $T=0.01$ and $1.00$~K. (c) First derivative of the
  differential conductance plotted in (b).\label{fig:goldModes_iets}}
\end{figure*}

So far we have just considered the energy-dependent transmission of the
elastic term, Eq.~(\ref{eq:Iel0}), of the total current in
Eq.~(\ref{eq:Itot}). Now we include additionally the effects due to the EV
coupling, as described by Eqs.~(\ref{eq:deltaIel0}) and (\ref{eq:Iinel}). More
precisely, we determine in the following the influence of vibrations on the
electric current in the WBL, using Eqs.~(\ref{eq:IelWBL})-(\ref{eq:IinelWBL}).

We have considered two different DRs, where we take the EV interaction into
account (see Fig.~\ref{fig:goldECC_tran}a).  In DR1 we include the EV coupling
just for the four chain atoms, in DR2 for all Au atoms in the C region. DR1
with the four dynamic atoms yields 12 vibrational modes, while the 12 dynamic
atoms in DR2 lead to 36 modes. Figure~\ref{fig:goldModes_iets}a shows all the
vibrations for DR1 (degenerate transversal modes are depicted only once),
while we have selected only those for DR2, which resemble the modes of
DR1. Despite the relaxation of all the C atoms, the symmetry of the ideal
contact is only slightly perturbed. Hence the transverse modes (V4-V6 and V8)
remain basically two-fold degenerate. Expanding the vibrationally active
region from DR1 to DR2 causes a blue-shift of the frequencies for the four
modes with the highest energy (V1-V4). For the other four modes (V5-V8), the
frequencies are red-shifted instead. Overall, however, the changes in the
frequencies remain relatively small.

\begin{figure*}[t]
 \sidecaption
 \centering{}\includegraphics[width=1.5\columnwidth]{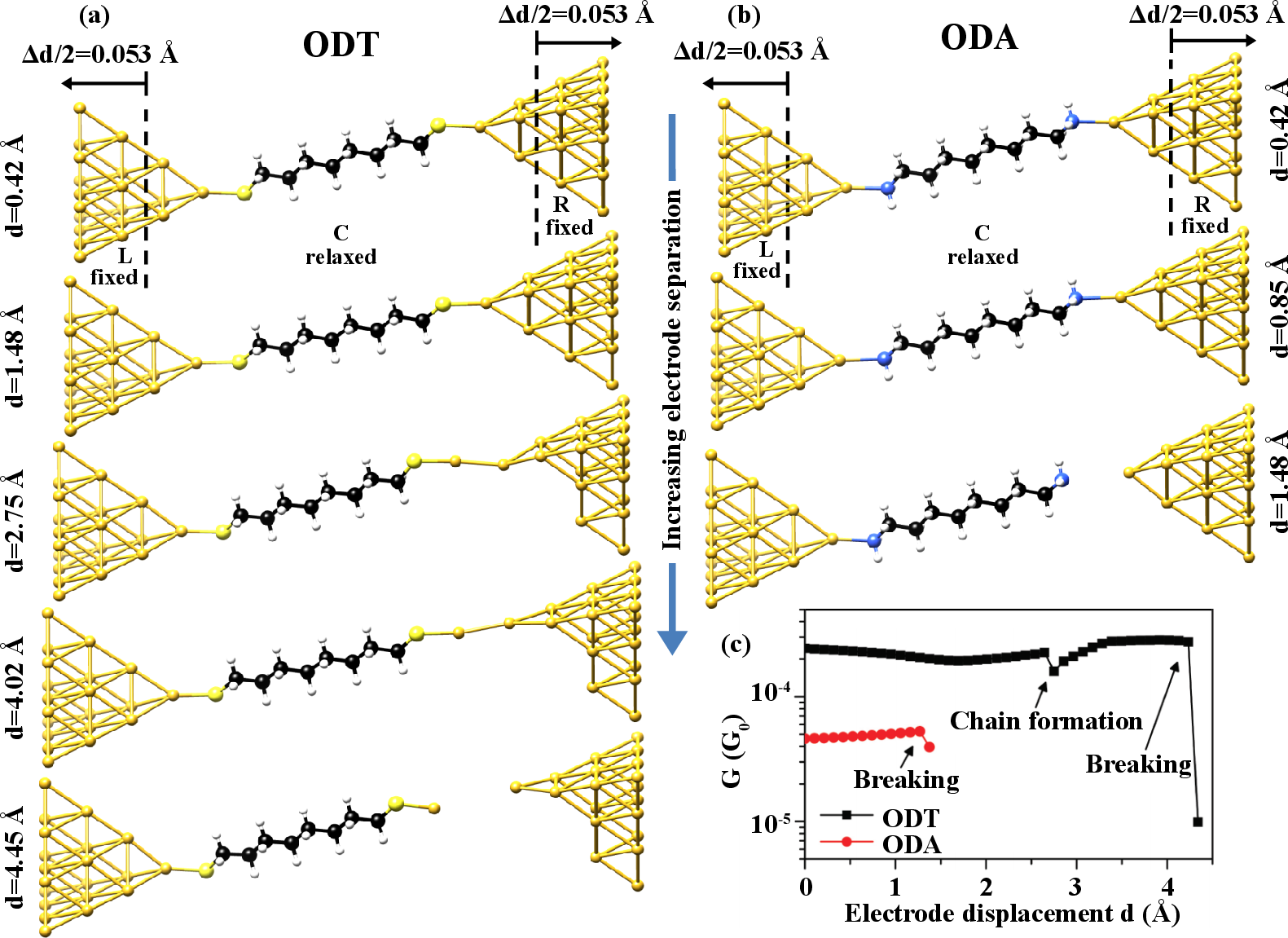}
\caption{Geometries for (a) ODT and (b) ODA molecules between Au electrodes at
  different electrode displacements. (c) Evolution of the elastic conductance
  as a function of the displacement.\label{fig:octacestretch}}
\end{figure*}

The calculated differential conductance as a function of voltage and its
derivative are shown in Fig.~\ref{fig:goldModes_iets}b and
\ref{fig:goldModes_iets}c for a vibrational broadening of $\eta=0.01$~meV.  We
observe that the three longitudinal modes (V1-V3) lead to the largest change
in the current. Since the longitudinal modes mainly couple to the first
transmission channel due to symmetry, they tend to decrease the conductance
consistent with the ``1/2 rule''
\cite{Paulsson2005,Vega2006,Paulsson2008}. The mode V1 gives rise to the
largest decrease of the conductance followed by V2 and V3. Comparing the
results for DR1 and DR2, we find that the curves remain very similar, but the
signals from V1 to V3 are shifted to higher bias voltages for DR2, as expected
from the frequency shifts in Fig.~\ref{fig:goldModes_iets}a. The additional
modes of DR2, which are mainly localized in the electrodes, do not give rise
to pronounced signals in the $\mbox{d}I/\mbox{d}V$. The increase in the
conductance at energies of the transversal mode V4 is due to its coupling to
the low-transmitting, $d$-like channels 3 and 4
\cite{Paulsson2005,Vega2006,Paulsson2008,Bohler2009}. Increasing the
temperature from $T=0.01$ to $1.00$~K tends to smear out the sharp steps in
the $\mbox{d}I/\mbox{d}V$.  All the presented results for DR1 are in good
agreement with previous tight-binding and ab-initio studies
\cite{Frederiksen2007b,Viljas2005}.

\subsection{Single-molecule gold-octane-gold junctions\label{sub:Octane}}

The electronic and vibrational properties of single-molecule junctions are
determined by the electrodes, the contacted molecule and the
molecule-electrode interfaces. They are reflected in the IET spectra, which
turn out to be a sensitive probe to characterize the junctions. This regards
for instance molecule-specific signatures to prove the presence of a
particular molecule between the electrodes or sensitivity to geometrical
changes during the junction elongation.

In Ref.~\cite{Kim2011} we analyzed both experimentally and theoretically
octanedithiol (ODT) and octanediamine (ODA) single-molecule junctions. The
focus was on the properties of the metal-molecule interface, and we showed
that the two different anchoring groups give rise to qualitatively different
features in the IET spectra. For sulfur anchors, which bind strongly to gold
surfaces, the $\text{S-Au}$ modes remained approximately constant in energy
during the junction elongation. Additionally, we observed IET peaks
corresponding to the formation of gold chains. For the much weaker
$\text{NH}_{2}\text{-Au}$ bond we found instead a red-shift of the
$\text{N-Au}$ mode with increasing electrode separation, and chain formation
was absent. Here we will extend this work and analyze theoretically in more
detail the IET signals related to vibrational modes localized on the molecule.

The ECC for the ODT and ODA junctions is shown in
Fig.~\ref{fig:octacestretch}a and \ref{fig:octacestretch}b. It is divided into
the L, C and R regions at the dashed lines indicated for the topmost
geometries. The two outermost layers of the Au electrodes are the L and R
parts. They are kept fixed at their ideal face-centered cubic lattice
positions, while the inner part or C region is relaxed to its ground-state
geometry. The vibrationally active region is identical to C. To simulate an
adiabatic stretching of the molecular junctions, we proceed as described in
Ref.~\cite{Kim2011}. The electrodes to the left and the right are separated
symmetrically by $\Delta d=0.2\:\text{a.u.}\approx 0.106\:\text{\angstrom}$ in each step. In
Fig.~\ref{fig:octacestretch}a and \ref{fig:octacestretch}b selected stages of
the stretching process are displayed, and the elastic conductance for each
electrode separation is summarized in Fig.~\ref{fig:octacestretch}c. For ODA
the dependence of the conductance on the electrode separation is rather weak,
showing a slight increase until the contact breaks at $d=1.48$~\angstrom. The
overall length of the conductance plateau for ODT is much longer than for ODA
and the conductance-distance trace exhibits several distinct features: At
first the conductance remains roughly constant before a kink appears at
$d=2.75$~\angstrom. It coincides with a plastic deformation of the contact,
resulting in the formation of a gold chain (see
Fig.~\ref{fig:octacestretch}a). Before the contact breaks at
$d=4.34$~\angstrom, the conductance increases with $d$ roughly to its starting
value at $d=0$. For a more detailed discussion we refer to our previous work
in Ref.~\cite{Kim2011}.

\begin{figure}
\centering{}\includegraphics[width=8.2cm]{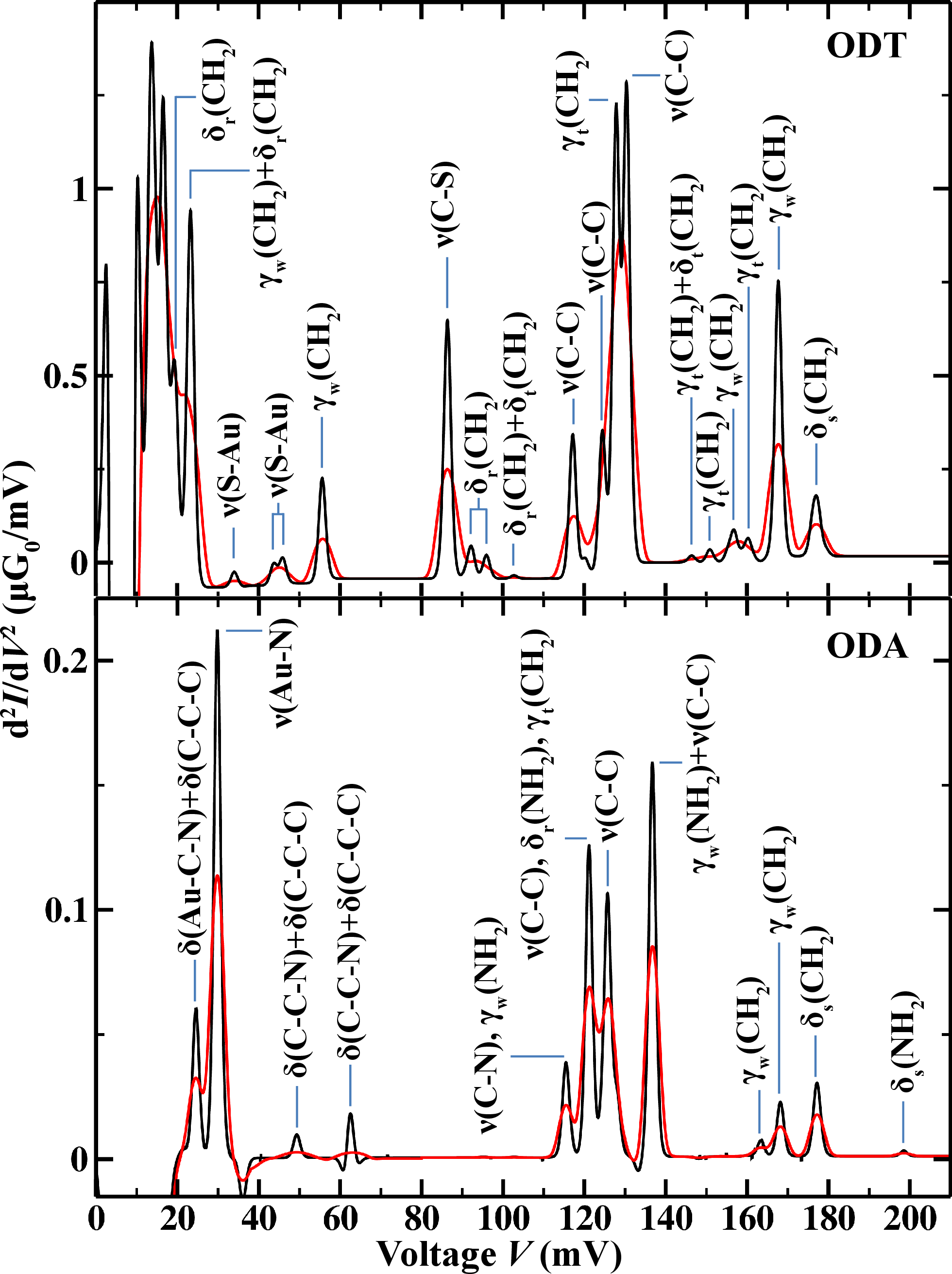}\caption{IET spectra
  without (black curve) and with (red curve) lock-in broadening for ODT and
  ODA. When we separate modes by a comma, there are several contributing to
  the same peak. When we use ``$+$'', a single mode has a mixed character.
\label{fig:ietsSNH2}}
\end{figure}

Next, we discuss inelastic effects in electrical transport due to the EV
coupling by considering the IET spectra of the ODT and ODA junctions. We use
the same terminology for the modes as in our Ref.~\cite{Kim2011} and refer the
interested reader to that work for its detailed description. We note that the
characterization of the vibrations in terms of molecule-internal modes remains
approximate due to the absence of molecular symmetries in the junctions. As
compared to Ref.~\cite{Kim2011}, we focus here mainly on the vibrational modes
of the octane molecules that do not involve the Au electrodes.

For all the IET spectra, calculated within the WBL, we assumed a temperature
of $T=4.2$~K and a vibrational broadening of $\eta=1$~meV. They describe the
intrinsic line-width broadening of the IET signals due to a finite temperature
and finite life-time of the vibrational modes, respectively. In the
experiments, the lock-in measurement technique constitutes another source of
broadening. It can be accounted for by convoluting the
$\mbox{d}^{2}I/\mbox{d}V^{2}$ with an instrumental function, which depends on
the modulation voltage $V_{\omega}$, applied to detect the IET characteristics
\cite{Klein1973,Hansma1977}. In Fig.~\ref{fig:ietsSNH2} the effect of this
broadening on the IET spectra is demonstrated for $V_{\omega}=5$~mV, the
modulation voltage used in Ref.~\cite{Kim2011}. The additional smearing may
prevent experimental resolution of individual vibrational modes, if they are
very close in energy. Even if the lock-in broadening is important for the
comparison of calculated IET spectra and measured ones, we neglect it in the
following calculations, as we did for the Au junctions above. In this way we
finely resolve all the vibrational features and consider only the intrinsic
broadening effects.

The electron-phonon coupling in ODT junctions has already been the subject of
several previous experimental
\cite{Wang2004,Arroyo2010,Kim2011,Lee2005,Okabayashi2008} and theoretical
\cite{Pecchia2004,Solomon2006,Paulsson2009} studies.  Our calculated
vibrational frequencies and IET spectra are consistent with these works. The
IET spectra for ODT and ODA, displayed in Fig.~\ref{fig:ietsSNH2}, are for a
stretching distance of $d=0.42$~\angstrom, where the molecules adopt a rather
straight configuration inside the junctions. At that $d$, the bond lengths
$d_{\text{C-C}}$ and bond angles $\alpha_{\text{C-C-C}}$ are very similar for
ODT and ODA.

When we compare the IET spectra at high energies from $145$ to $200$~meV, we
find that the positions of the pronounced peaks stemming from
$\gamma_{\text{w}}(\text{CH}_{2})$ and $\delta_{\text{s}}(\text{CH}_{2})$
modes, located at 169 and 178 meV, respectively, are the same for ODT and
ODA. The position of the second $\gamma_{\text{w}}(\text{CH}_{2})$ mode at 156
meV for ODT and 163 meV for ODA differs slightly. For ODT we observe small
signals from $\gamma_{\text{t}}(\text{CH}_{2})$ modes at 150 and 160 meV as
well as a
$\gamma_{\text{t}}(\text{CH}_{2})\mbox{+}\delta_{\text{t}}(\text{CH}_{2})$
mode at 145 meV, which are absent for ODA. The very faint signal at $197$~meV
can be attributed to a $\delta_{\text{s}}(\text{NH}_{2})$ vibration of the ODA
anchoring group.  Between $110$ and $145$~meV, $\nu(\text{C-C})$ modes give
rise to prominent features in the IET spectra. We can identify three major
peaks at slightly different energies. For ODT the $\nu{(\text{C-C})}$ mode
with the highest energy is located at $131$~meV, while it is slightly
blue-shifted to $137$~meV for ODA and contains an additional
$\gamma_{\text{w}}(\text{NH}_{2})$ contribution. At $125$~meV we observe a
clear $\nu{(\text{C-C})}$ mode for both anchoring groups.  The third
$\nu{(\text{C-C})}$ mode is located at $118$~meV for ODT and at a slightly
higher energy of $121$~meV for ODA. While the $\nu{(\text{C-C})}$ modes with
the highest and lowest energies are blue-shifted for ODA as compared to ODT,
the $\gamma_{\text{t}}(\text{CH}_{2})$ mode is red-shifted and located at
$128$~meV for ODT and $121$~meV for ODA. Not considering modes with a mixed
character, we find three signals involving the N atom of the amino anchoring
groups in the energy range: One $\delta_{\text{r}}(\text{NH}_{2})$ mode at
$121$~meV and two modes $\nu(\text{C-N})$, $\gamma_{\text{w}}(\text{NH}_{2})$
at $116$~meV. Between $70$ and $110$~meV there are no signatures of
vibrational modes present in the IET spectra for ODA. ODT on the other hand
shows some small signals belonging to $\delta_{\text{r}}(\text{CH}_{2})$ and
$\delta_{\text{t}}(\text{CH}_{2})$ vibrations between $90$ and $105$~meV. The
magnitude of these two IET signals is known to depend crucially on the precise
contact geometry \cite{Solomon2006}. In contrast, the $\nu(\text{C-S})$ mode
at $86$~meV gives rise to a dominant peak. At energies below $70$~meV, the IET
spectra of ODT and ODA differ substantially. In that range we observe most of
the modes involving the anchoring groups and the Au electrodes. Additionally,
we observe several $\delta(\text{C-C-C})$ and $\delta(\text{C-C-N})$
stretching modes for ODA and some low-energy $\gamma_\text{w}(\text{CH}_{2})$
and $\delta_\text{r}(\text{CH}_{2})$ modes for ODT. The modes at very low
energies are mainly localized on the Au electrodes and shall not be discussed
here. Our results confirm that the characteristic peaks in the IET spectra and
their sensitivity to the precise contact geometry can be used to infer the
precise binding geometry and to distinguish between ODT and ODA,
i.e.\ molecules that differ only in their anchoring group \cite{Kim2011}.

\begin{figure}
\centering{}\includegraphics[width=8.4cm]{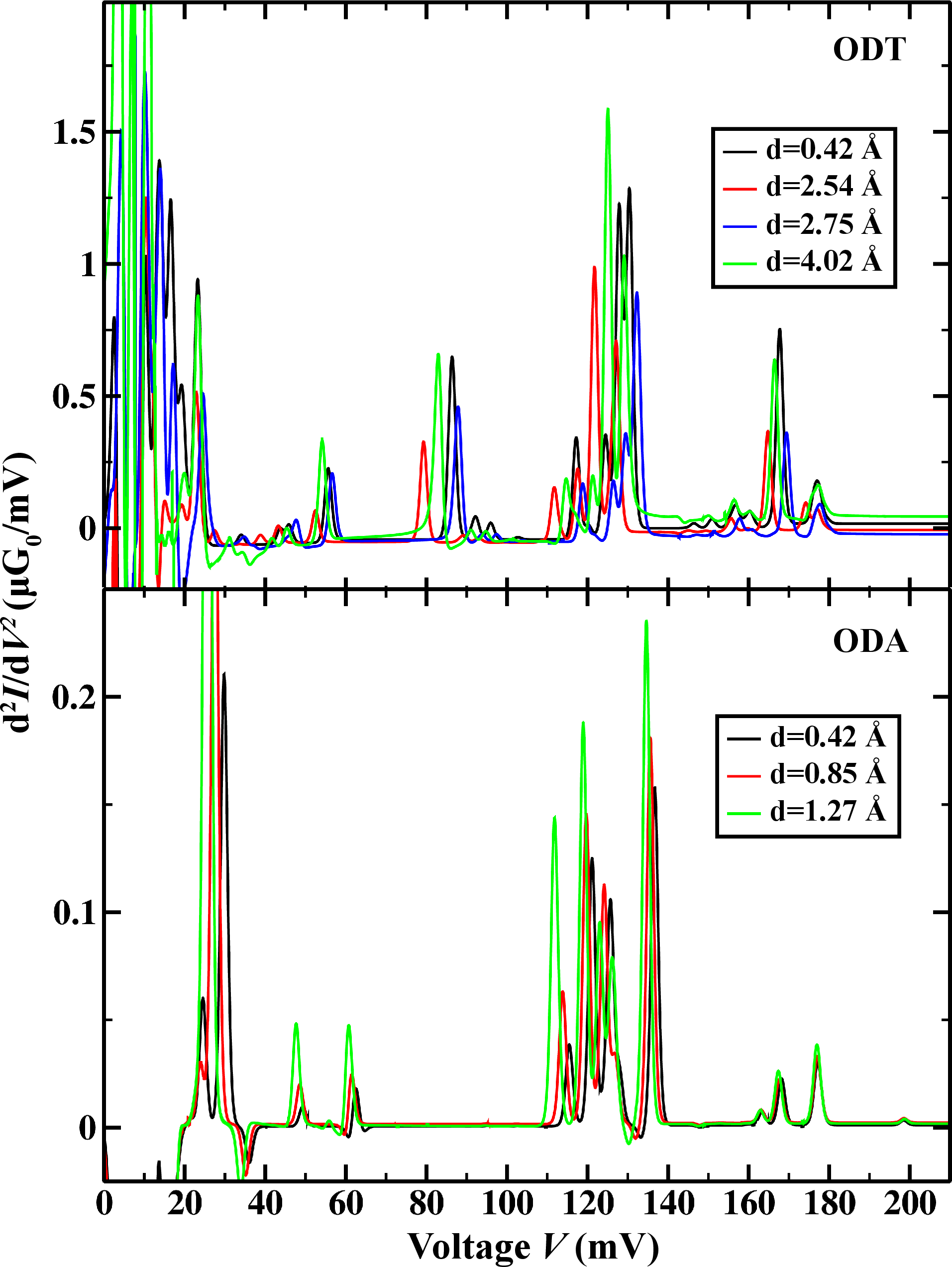}\caption{IET spectra
  for different electrode displacements of the ODT and ODA
  junctions.\label{fig:iets-odt-oda-stretch}}
\end{figure}

The evolution of the IET spectra with increasing electrode separation is
displayed in Fig.~\ref{fig:iets-odt-oda-stretch} for both ODT and
ODA. Increasing $d$, increases $d_{\text{C-C}}$ and $\alpha_{\text{C-C-C}}$ in
the octane molecule. This affects mainly vibrational modes involving the
carbon backbone. For ODA the $\delta(\text{C-C-C})$, $\delta(\text{C-C-N})$
and $\nu(\text{C-C})$ modes are red-shifted by around $3$~meV. The red-shifts
occurs continuously during the stretching process from $d=0$ until the
contract breaks at $d=1.38$~\angstrom. The energy of the modes involving the
${\text{NH}_{2}}$ anchoring group [$\delta_{s}(\text{NH}_{2})$] and the
$\text{CH}_{2}$ units [$\gamma_{\text{w}}(\text{CH}_{2})$,
  $\delta_{\text{s}}(\text{CH}_{2})$] remains basically constant. As expected,
they are not influenced by the mechanical stretching. For ODT a similar
behavior is observed, but the plastic deformation of the Au electrodes needs
to be taken into account. From $d=0$ to $d=2.54$~\angstrom\ the ODT contact is
elastically deformed, building up stress in the molecule. As for ODA,
$d_{\text{C-C}}$ and $\alpha_{\text{C-C-C}}$ are increased, resulting in a
red-shift of $\nu(\text{C-C})$ modes by around $7$ to $9$~meV. For the modes
with clear $\text{CH}_{2}$ character, the red-shift is smaller and does not
exceed $3$~meV. The plastic deformation of the Au electrode, occurring at
around $d=2.75$ \angstrom, releases the stress and allows $d_{\text{C-C}}$ and
$\alpha_{\text{C-C-C}}$ to restore their initial values. This results in a
blue-shift of the vibrational modes which take values slightly above their
initial energy. The movement of the peaks in the IET spectra in the subsequent
elastic stage until contact rupture is similar to the first one.

\section{Conclusions\label{sec:Conclusions}}

We presented a new first-principles approach to study the IET spectra of
atomic and molecular contacts. To achieve this, we extended the quantum
chemistry software package TURBOMOLE to compute the EV couplings via an
efficient and accurate semi-analytical derivative scheme based on DFPT. This
functionality will be available in TURBOMOLE 6.6. It allows us to describe the
coupled system of electrons and phonons of the nanostructures at the level of
DFT without free parameters. Using a LOE in terms of the EV coupling
\cite{Viljas2005}, we determined the influence of vibrations on the electrical
current. This constitutes an important extension of our previous capabilities
to study the elastic transport properties of nanoscale conductors
\cite{Pauly2008}. Gold electrodes bridged by an atomic chain served as test
system and demonstrated that our approach is consistent with experimental and
theoretical studies in the literature.

Based on these theoretical developments, we studied the IET spectra of ODT and
ODA single-molecule junctions. We extended our previous investigations in
Ref.~\cite{Kim2011}, where we focused on the metal-molecule interface, by a
more detailed discussion of the molecular vibrations localized on the octane
itself. We found that the vibrations of the alkane backbone differ only
slightly for the two different anchoring groups. However,
anchoring-group-specific modes could be clearly resolved in the IET spectra,
offering the possibility to distinguish between ODT and ODA. The sulfur and
amine anchors differ substantially in their binding strength to Au. This
resulted in a qualitatively different behavior of the junctions during their
elongation. While the ODA junction broke at the weak Au-N bond, gold
electrodes were plastically deformed for ODT. Ultimately, the ODT junction
ruptured at a Au-Au bond and not the strong Au-S bond. During the junction
elongation we observed a red-shift of $\delta(\text{C-C-C})$ and
$\nu(\text{C-C})$ modes for both anchoring groups due to the increasing bond
lengths $d_{\text{C-C}}$ and bond angles $\alpha_{\text{C-C-C}}$. The
methylene vibrations on the other hand were only slightly affected by the
increasing electrode separation. For ODT the plastic deformation of the Au
electrodes was reflected also in the molecular vibrations.  Due to the
relaxation of the stress built up during elastic stages, the plastic
deformation lead to a decrease of $d_{\text{C-C}}$ and
$\alpha_{\text{C-C-C}}$, blue-shifting the corresponding vibrational
modes. Subsequent elastic deformations decreased their energies again.

So far, heating effects and the non-equilibrium distribution of the phonons
are taken into account in an approximate way in our ab-initio calculations.
The consideration of the coupling of the vibrations in the central device
region to those in the electrodes \cite{Romano2007,Engelund2009} constitutes a
natural extension of the developed methodology. Work along these lines
combined with calculations for a larger set of atomic and molecular junctions
will allow new insights into the interaction of electrons and phonons,
important for challenges such as the improved control of electron transport
and heating in electronic nanocircuits.

\begin{acknowledgement}
We thank Y.\ Kim for his contributions to the experimental work in
Ref.~\cite{Kim2011}, J.\ C.\ Cuevas for sti\-mu\-la\-ting discussions, and the
TURBOMOLE GmbH for providing us with the source code of TURBOMOLE.  The work
of M.B.\ was supported through the DFG priority program 1243 and a FY2012
(P12501) Postdoctoral Fellowship for Foreign Researchers from the Japan
Society for Promotion of Science (JSPS) as well as by a JSPS KAKENHI, i.e.,
``Grant-in-Aid for JSPS Fellows'', Grant No.\ 24$\cdot$02501. In addition,
J.K.V.\ gratefully acknowledges funding through the Academy of Finland,
T.J.H.\ through the Baden-W\"urttemberg Stiftung within the Research Network
of Excellence ``Functional Nanostructures'', E.S.\ through the DFG via SPP
1243 and SFB 767, G.S.\ through the Initial Training Network ``NanoCTM''
(Grant No.\ FP7-PEOPLE-ITN-2008-234970), and F.P.\ through the DFG Center for
Functional Nanostructures (Project C3.6) and the Carl Zeiss foundation.
\end{acknowledgement}

%
%

\providecommand{\WileyBibTextsc}{}
\let\textsc\WileyBibTextsc
\providecommand{\othercit}{}
\providecommand{\jr}[1]{#1}
\providecommand{\etal}{~et~al.}

\end{document}